\documentclass[twocolumn]{aastex631}

\newcommand{\sulfurdioxide}{SO$_2$}
\newcommand{\um}{$\mu$m}
\usepackage{amsmath,bm}
\usepackage{savesym}
\begin{document}

\title{Potential Melting of Extrasolar Planets by Tidal Dissipation}
\shorttitle{Volcanism on Exoplanets}
\shortauthors{Seligman et al.}

\author[0000-0002-0726-6480]{Darryl Z. Seligman}
\altaffiliation{NSF Astronomy and Astrophysics Postdoctoral Fellow}
\affiliation{Department of Astronomy and Carl Sagan Institute, Cornell University, 122 Sciences Drive, Ithaca, NY, 14853, USA}

\correspondingauthor{Darryl Z. Seligman}
\email{dzs9@cornell.edu}

\author[0000-0002-9464-8101]{Adina D. Feinstein}
\altaffiliation{NHFP Sagan Fellow}
\affiliation{Department of Astronomy and Astrophysics, University of Chicago, Chicago, IL 60637, USA}
\affiliation{Laboratory for Atmospheric and Space Physics, University of Colorado Boulder, UCB 600, Boulder, CO 80309}

\author[0000-0002-1934-6250]{Dong Lai}
\affiliation{Department of Astronomy and Carl Sagan Institute, Cornell University, 122 Sciences Drive, Ithaca, NY, 14853, USA}

\author[0000-0003-0156-4564]{Luis Welbanks}
\affiliation{School of Earth \& Space Exploration, Arizona State University, Tempe, AZ, 85257, USA}

\author[0000-0002-0140-4475]{Aster G. Taylor}
\altaffiliation{Fannie and John Hertz Foundation Fellow}
\affiliation{Dept. of Astronomy and Astrophysics, University of Chicago, Chicago, IL 60637, USA}
\affiliation{Dept. of Astronomy, University of Michigan, Ann Arbor, MI 48109}

\author[0000-0002-7733-4522]{Juliette Becker}
\affiliation{Division of Geological and Planetary Sciences, California Institute of Technology, Pasadena, CA 91125}
\affiliation{Department of Astronomy, University of Wisconsin-Madison, 475 N. Charter St., Madison, WI 53706, USA}
\author[0000-0002-8167-1767]{Fred~C.~Adams}
\affiliation{Physics Department, University of Michigan, Ann Arbor, MI 48109}
\affiliation{Astronomy Department, University of Michigan, Ann Arbor, MI 48109}
\author[0000-0003-4022-6234]{Marvin Morgan}
\affiliation{Department of Astronomy, The University of Texas at Austin, Austin, TX 78712, USA}
 \author[0000-0002-8716-0482]{Jennifer B. Bergner}
\affiliation{Department of Chemistry, University of California, Berkeley, Berkeley, CA, 94720}

\begin{abstract}
 Tidal heating on Io  due to its finite eccentricity was predicted to drive surface volcanic activity, which was subsequently confirmed  by the  \textit{Voyager} spacecrafts.  Although the volcanic activity in Io is more complex, in theory volcanism can be driven by runaway melting in which the  tidal heating increases as the mantle thickness decreases.   We show that this runaway melting mechanism  is  generic for a composite planetary body with liquid core and solid mantle, provided that (i) the mantle rigidity, $\mu$, is comparable to the central pressure, i.e. $\mu/ (\rho g R_{\rm P})\gtrsim0.1$ for a body with density  $\rho$, surface gravitational acceleration $g$,  and radius $R_{\rm P}$, (ii) the surface is not molten, (iii) tides deposit  sufficient energy, and (iv) the planet has nonzero eccentricity.  We calculate the  approximate liquid core radius as a function of $\mu/ (\rho g R_{\rm P})$, and find that more than $90\%$ of the core will melt due to this runaway for $\mu/ (\rho g R_{\rm P})\gtrsim1$.   From all currently confirmed exoplanets, we find that the terrestrial planets in the L98-59 system are the most promising candidates for sustaining active volcanism. However,  uncertainties regarding the quality factors and the details of tidal heating and cooling mechanisms prohibit definitive claims of volcanism on any of these planets. We generate synthetic transmission spectra of these planets assuming Venus-like atmospheric compositions with an additional 5, 50, and $98\%$ SO$_2$ component, which is a tracer of volcanic activity. We find a $\gtrsim 3 \sigma$ preference for a model with SO$_2$ with 5-10 transits with \textit{JWST} for L98-59bcd. 
\end{abstract}


\section{Introduction} \label{sec:intro}

\citet{Peale1979}  predicted that the \textit{Voyager}  flybys would observe widespread volcanic activity on the surface of Io. This volcanic activity is driven by tidal heating on Io as a result of its eccentric orbit around Jupiter --- this finite eccentricity is forced by a Laplace resonance between Io, Europa and Ganymede. In particular, \citet{Peale1979} described a runway melting process in which  the tidal heating rate increases with decreasing width of the solid mantle. Active volcanism was subsequently observed in \textit{Voyager} spacecraft images \citep{Smith1979,Morabito1979,Strom1979} and later by the 
 \textit{Galileo} \citep{McEwen1998,Spencer2000} and \textit{New Horizons} spacecrafts \citep{Spencer2007}. However, this melting mechanism predicts the presence of a very thin mantle which would not support the mountainous topography of Io. More sophisticated models have attributed the volcanism from tidal heating to partial melting \citep{Lopes2007} or a subsurface magma ocean \citep[for a review of interior models see][]{Keane2023}.
 
\textit{Pioneer 10} measurements of the   ionosphere provided  evidence that the  atmosphere was enriched with sulfur dioxide (SO$_2$) injected by volcanic activity   \citep{Kliore1974,Kliore1975}. In addition, \textit{Voyager} ultraviolet (UV) observations of the surrounding space environment \citep{Broadfoot1979,Bridge1979} revealed the presence of ionized sulphur and oxygen atoms in what is now referred to as the Jovian ``plasma torus'' \citep[also see][]{Moos1991,Thomas1992,Thomas1996,Steffl2004,Murakami2016,Volwerk2018}.

While Io is the most volcanically active body within the Solar System, lesser degrees of current or recent  volcanism have been  inferred in several other locations \citep{Wilson1983}. For example, \citet{Solomon1982} hypothesized that previous generations of volcanism could explain volcanic deposits observed on the  Venusian terrain.  Recently, \citet{herrick_hensley_2023}  confirmed that the Atla Regio region containing two of the largest Venusian volcanoes --- Ozza Mons and Maat Mons ---   was volcanically active. The evidence of past volcanism on Mars has been apparent from in situ  analyses of Martian soil  by the Viking lander \citep[e.g.][]{Greeley2000}.  The MESSENGER spacecraft \citep{Solomon2007} imaged the surface of Mercury and provided evidence for previous, extended volcanism \citep{Head2008,Head2009,Denevi2009,Prockter2010}. Although it is not active today, the Earth's Moon itself  also used to be volcanically active \citep{Greeley1971,Spudis2013}. 

The evidence outlined above from our Solar System indicates that a wide variety of environments and conditions are amenable  to planetary volcanism. In this paper, we consider  the applicability of the  runaway melting mechanism to extrasolar planets (exoplanets). In a series of papers, \citet{Jackson2008a} and \citet{Barnes2009,Barnes2009b} outlined the properties of exoplanets that are required to produce volcanism driven by tidal and internal heating, with a particular focus on the ramifications for planetary habitability.  This analysis was extended by \citet{Quick2020}, who identified 53 exoplanets with properties amenable to surface or cryogenic volcanism.  \citet{Barr2018} argued that Trappist-1 b and c may maintain  magma oceans in their rocky mantles and that the latter may have surface eruptions of silicate magma  detectable with \textit{JWST} via calculations of the  interior structure and temperature induced by tidal dissipation (see also \citet{Dobos2019}).

A distinct class of exoplanets, known as ``lava worlds,'' are planets with sufficiently high effective temperatures to maintain entirely molten surfaces  \citep{Leger2009,Batalha2011,Winn2018}. While no lava worlds exist within the Solar System, terrestrial and sample return geochemical evidence indicates that silicate mantles of many Solar System bodies underwent molten phases \citep{ElkinsTanton2012}. \textit{Spitzer Space Telescope} observations of  55 Cancre e revealed a day-side temperature of $2700\pm270$ K sufficient to entirely melt the surface \citep{Demory2016}. Moreover, these  observations  were consistent with low viscosity magma flows on the surface.  \citet{RiddenHarper2016}  reported a tentative detection of a sodium feature in the same planet that may also be associated with active volcanism. For recent reviews of lava worlds, we guide the reader to \citet{Chao2021}. While these planets are relevant, we focus in this paper on planets with lower surface temperatures but sufficient tidal heating to power volcanism. 

There have also been investigations into the possibility that Jupiter-Io analogues could be detected in exoplanetary-exomoon systems via tidally induced volcanic activity. \citet{Oza2019} argued that Na I and K I surrounding gas giant exoplanets could be the signature of volcanic activity of an Io-like satellite. This assertion was motivated by the Na I and K I clouds that have been observed on Io \citep{Brown1974,Brown1974b,Trafton1975}.  These authors demonstrated that the high-altitude Na signature of WASP 49-b \citep{Wyttenbach2017} could be explained by an outgassing Io-like satellite. \citet{Rovira_Navarro2021} investigated the extent to which tidally heated and potentially volcanic exomoons would be detectable via direct imaging \citep[see also][]{Tokadjian2023}.  \citet{kleisioti2023tidally}  considered the detectability  of an exomoon around $\varepsilon$ Eridani b via volcanic activity with \textit{JWST}. 

In this paper, we focus on  rocky planets around low-mass stars with orbits capable of producing volcanism via the melting mechanism. Specifically, we focus on the applicability of the runaway mechanism to produce surface volcanism on a rocky exoplanet. In \S \ref{sec:melting} we generalize the runaway melting mechanism to include larger bodies. In \S \ref{sec:exocandidates} we identify currently known exoplanets that experience tidal heating comparable to or greater than Io (per unit mass) despite the uncertainty of the tidal parameters. In \S \ref{sec:future_prospect} we demonstrate that volcanic activity could be detected via SO$_2$ spectral features with \textit{JWST} on the L98-59 planets with a reasonable number of transits. We conclude in \S \ref{sec:conclusions} with a summary of our results and discussion of their implications. 

\begin{figure}[t]
    \centering
    \includegraphics[width=\linewidth]{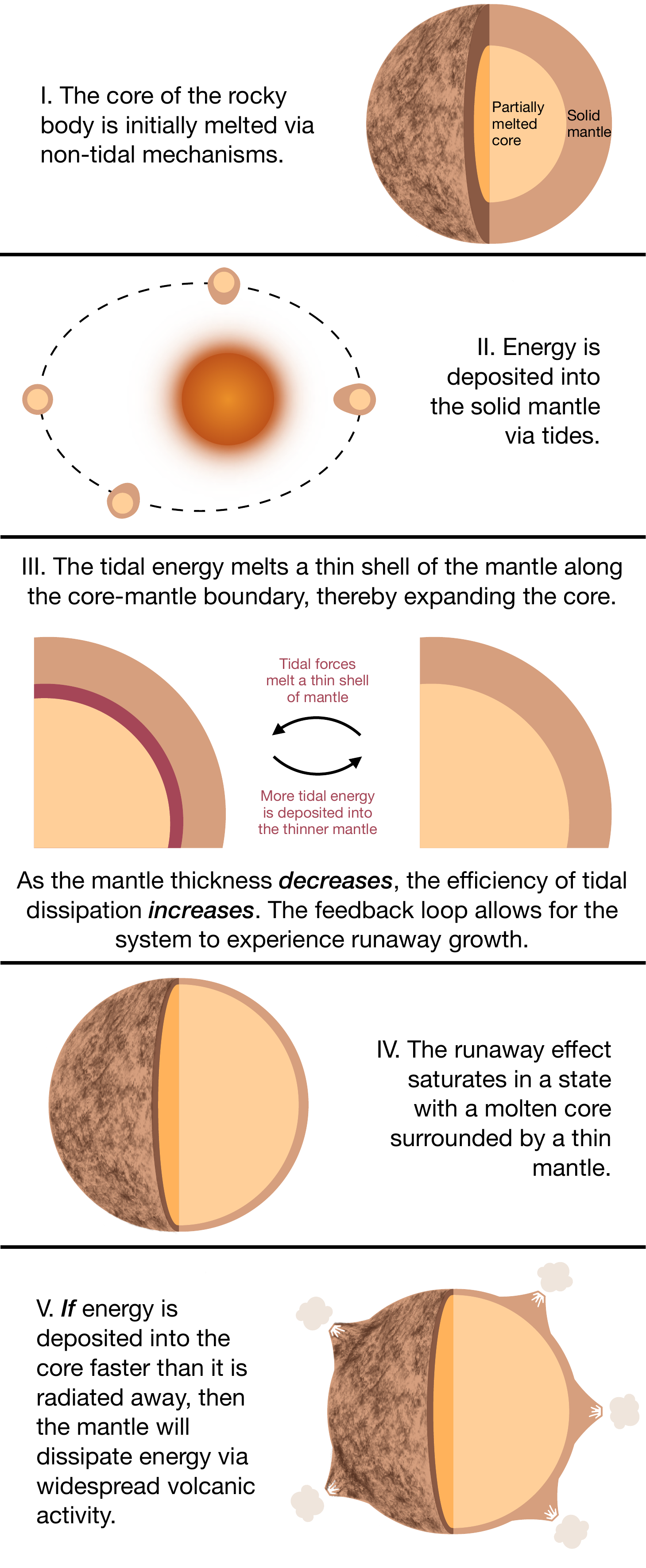}
    \caption{The mechanism in brief. A summary of runaway  melting process outlined by \citet{Peale1978}. This mechanism only operate if sufficient energy is deposited to raise the temperature above solidus.} 
    \label{fig:schematic}
\end{figure}

\section{Runaway Melting}\label{sec:melting}

In this section, we describe the melting process described by \citet{Peale1978} and    \citet{Peale1979}. We use the derived results and notation from \citet{Beuthe2013}.   

\subsection{The Mechanism in Brief}\label{subsec:brief}

We consider a composite planet with a liquid core (for example melted via radioactive heating) and a rocky mantle. The heart of the mechanism stems from the fact that the volume-integrated deposition of energy from tidal stresses  \textit{increases} as the  mantle thickness \textit{decreases}.  This feedback mechanism allows for a runaway process: as a thin shell surrounding the core melts, more total energy is deposited in the now-thinner mantle, which leads to further melting --- as long as sufficient energy is deposited to raise the temperature above  solidus. The feedback loop leads to saturation when melting of the mantle no longer increases the efficiency of tidal dissipation.  As long as the tides are strong enough to melt the mantle faster than the body can radiate away the energy deposited, some of the tidally dissipated energy is released via widespread surface volcanism. We outline this process in Figure \ref{fig:schematic}.

\subsection{Details of the Melting Mechanism}\label{subsec:details}

The deformation of a two layer body with a melted core and a solid mantle by an external potential was first calculated by \citet{herglotz_1911} and updated by \citet{Harrison1963}. \citet{Zschau1978} extended the analysis to parameterize the total tidal heating in a stratified and compressible body in the imaginary part of the Love number, $\Im(k_2)$.  This analysis was extended by  \citet{Peale1978}, who first outlined the runaway melting mechanism with applications to Io \citep{Peale1979}. It should be noted that volcanism on Io has since been attributed to subtler effects such as  partial melting \citep{Lopes2007,Keane2023}. 

\citet{Beuthe2013} derived the Love number, $k_2$, for a composite planet consisting of a liquid core and a rocky mantle (with rigidity $\mu$), assuming the core and mantle have the same density, $\rho$. The result is

\begin{equation}\label{eq:k2_2layer}
    k_{2} = \frac{3}{2}\bigg(1+\mathcal{Z}(\eta) \frac{\mu}{\rho g R_{\rm P}}\bigg)^{-1}\,.
\end{equation}
In Equation (\ref{eq:k2_2layer}), the parameter $\eta=R_{\rm C}/R_{\rm P}$ is the liquid core radius, where $R_{\rm C}$ is the radius of the core, $R_{\rm P}$ is the total radius of the planet, and $g$ is the surface gravitational acceleration. The function $\mathcal{Z}(\eta) $ is derived in Equation (112) of \citet{Beuthe2013},

\begin{equation}\label{eq:zliq}
    \mathcal{Z}(\eta) = 12 \,\bigg(\,\frac{19-75\eta^3+112\eta^5-75\eta^7+19\eta^{10}}{24+40\eta^3-45\eta^7-19\eta^{10}}\,\bigg)\,.
\end{equation}

A viscoelastic mantle can be modeled by a complex rigidity $\tilde{\mu}$. The modulus of the complex number gives the amplitude and the argument gives the phase lag between the perturbation force and the resulting deformation. For the remainder of this work, a quantity with a tilde indicates that it has both a real and imaginary component.  Following the correspondence principle \citep{Biot1954}, we can obtain the complex Love number $\tilde{k}_2$ by replacing the real rigidity of an elastic solid with the complex rigidity $\tilde{\mu}$ of the viscoelastic solid. The complex rigidity $\tilde{\mu}$ is highly uncertain, and depends on the rheology of the mantle.

\begin{figure}[t]
    \centering
    \includegraphics[width=1\linewidth]{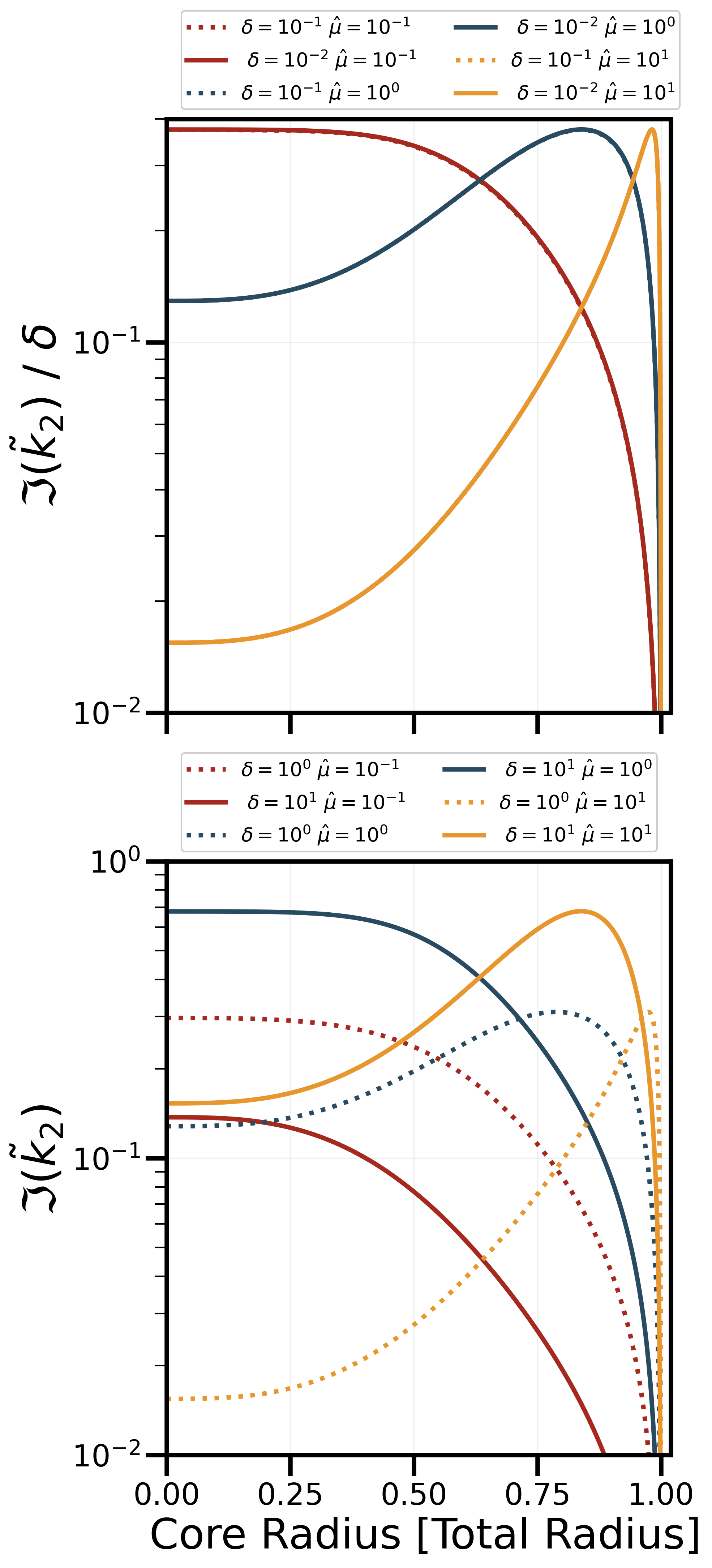}
    \caption{The  imaginary part of the Love number $\tilde{k}_2$ as a function of the liquid core radius, given by Equation (\ref{eq:finalk2}).  We show this a range of  $\hat{\mu}$ and $\delta$.  The upper and lower panels have different normalizations to compare the functions. }  
    \label{fig:imk}
\end{figure}
\begin{figure*}
    \centering
    \includegraphics[width=1\linewidth]{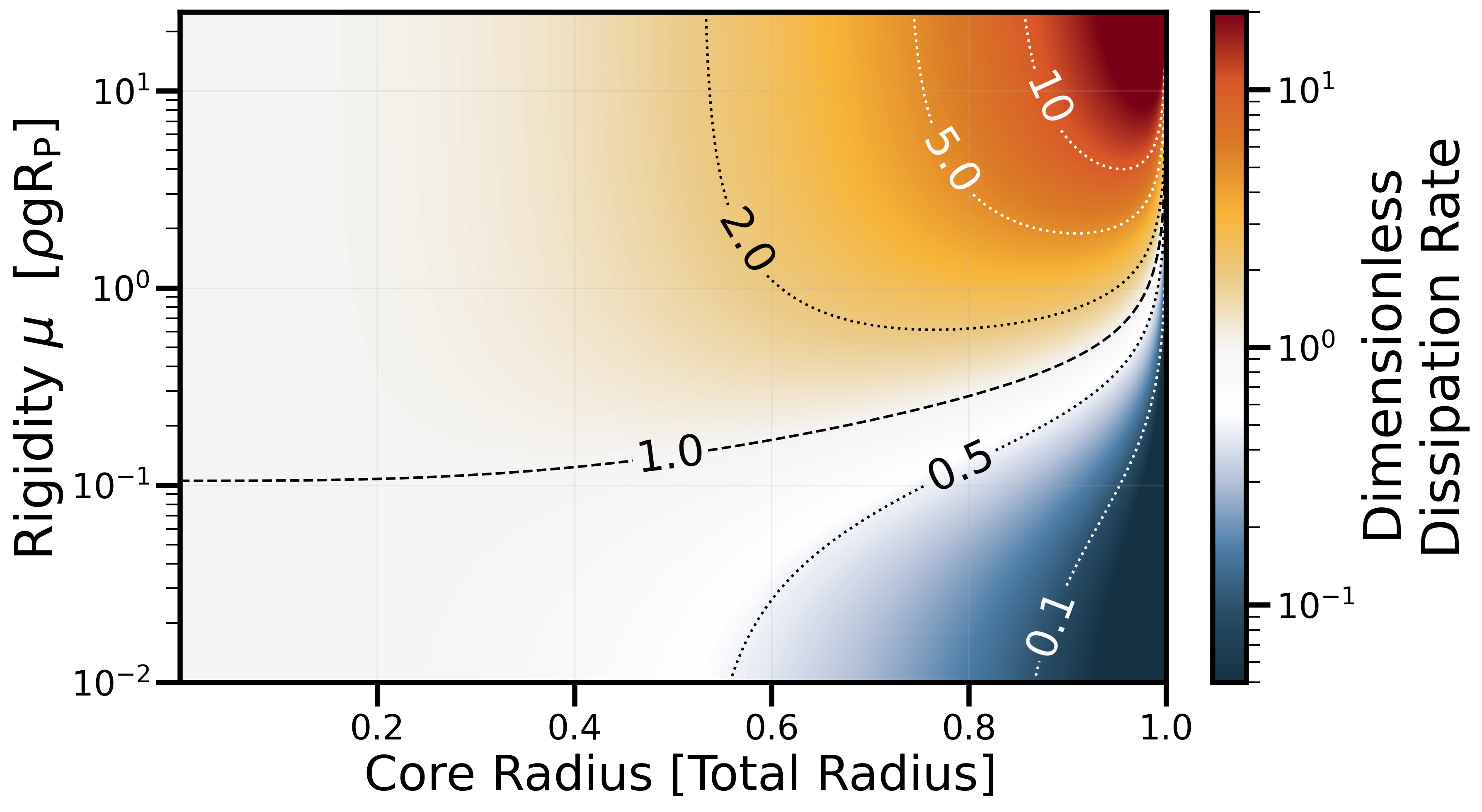}
    \caption{The total tidal heating for a planet with a melted core normalized by the total heating for the case of no core. The x-axis shows the liquid core radius $\eta=R_{\rm C} / R_{\rm P} $. The y-axis shows the dimensionless rigidity of the mantle $\hat{\mu}=\mu/\rho g R_{\rm P} $. The color scale indicates the total heating normalized by the  heating in the absence of a liquid  core, $\mathcal{E}(\eta)=\dot{E}_{\rm Heat}(\eta)\dot{E}_{\rm Heat}(\eta=0)$ (Equation (\ref{eq:epsilon})).  Contours corresponding to specific values of $\mathcal{E}(\eta)$ are indicated. } 
    \label{fig:instability}
\end{figure*}

\begin{figure*}
    \centering
    \includegraphics[width=1\linewidth]{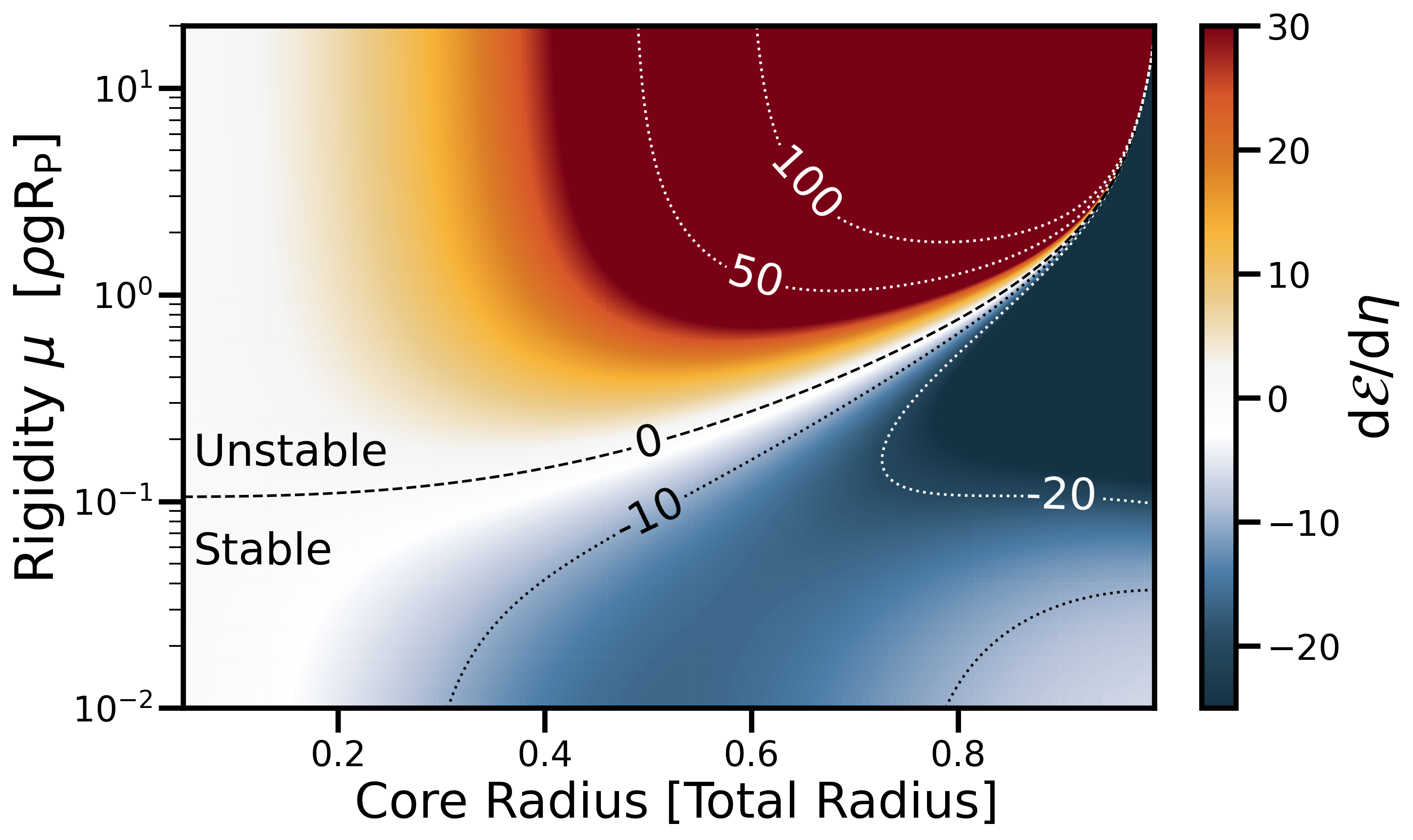}
    \caption{The stability of the   melting mechanism given by the derivative of the normalized total heating with respect to the liquid core radius. The color scale  shows the  value of the derivative of the heating ratio between the melted and homogeneous case. When this quantity is positive/negative the melting mechanism is unstable/stable as indicated in the annotations. The curve where the derivative is zero indicates the liquid core radius where  the mechanism saturates and corresponds to the approximate, idealized final core radius under the model assumptions.  } 
    \label{fig:stability_analysis}
\end{figure*}

The tidal heating rate, $\dot{E}_{\rm Heat}$, is related to $\Im(\tilde{k}_2)$ by \footnote{Note that for the complex ${\tilde k}_2$, we follow the sign convention
of \citet{Ogilvie2014} where the tidal forcing potential of frequency $\omega$ has the form
$\exp(-i\omega t)$. }
\begin{equation}\label{eq:tidalheating_1layer}
     \dot{E}_{\rm Heat} = 5 \Omega_{\rm P} \bigg(\frac{ G \, M_*^2\,R_{\rm P}^5}{ a^6}\,\bigg)\,\Im(\tilde{k}_2) \,\Psi_0 \big(e,I,\Delta \Omega/\Omega_{\rm P}\big)\,.
\end{equation}
In Equation (\ref{eq:tidalheating_1layer}), $\Omega_{\rm P}$ is the rotation rate of the planet, $M_*$ is the stellar mass, and $\Psi_0 (e,I,\Delta \Omega/\Omega_{\rm P}) $ is a dimensionless potential function of eccentricity $e$, obliquity $I$, and satellite rotation compared to the mean motion. For the case of zero obliquity ($I=0$) and synchronous rotation ($\Delta \Omega=\Omega_{\rm P}$), the dimensionless potential function is given by  $\Psi_0=21 e^2/5$. For this case the heating rate reduces to the form

\begin{equation}\label{eq:simpleheating}
\dot{E}_{\rm Heat}= \Omega_{\rm P} \bigg(\frac{ G \, M_*^2\,R_{\rm P}^5}{ a^6}\,\bigg)\,\Im(\tilde{k}_2)\bigg( \frac{21}{2}e^2\bigg)\,.
\end{equation}
Note that the conventional tidal Q  is defined as

\begin{equation}\label{eq:complex_love}
    \Im(\tilde{k}_2)= |\tilde{k}_2|\bigg/Q\,,
\end{equation}
where by construction in most cases $\Re(\tilde{k}_2)\gg\Im(\tilde{k}_2)$. Consider a simple model for $\tilde{\mu}$ of the form
\begin{equation}\label{eq:maxwell}
    \tilde{\mu} = \frac{\mu}{1+i \delta}\,,
\end{equation}
where $\mu$ and $\delta$ are real. For example, a Maxwell rheology has $\delta=\omega_M/\omega$ with $\omega_M$ the relaxation (Maxwell) frequency and the tidal frequency $\omega$. This ansatz implies that
\begin{equation}\label{eq:finalk2}
    \Im(\tilde{k}_2) = \frac{3}{2}\frac{\mathcal{Z}(\eta)\delta\hat{\mu} }{|1+\mathcal{Z}(\eta)\hat{\mu}|^2 +\delta^2}\,,
\end{equation}
where the quantity $\hat{\mu}$ is defined as 
\begin{equation}
    \hat{\mu}=\frac{\mu}{\rho g R_{\rm P}}\,.
\end{equation}

The corresponding $Q$-value is given by 
\begin{equation}
    \frac{1}{Q} = \frac{\mathcal{Z}(\eta)\delta\hat{\mu}}{\sqrt{(\mathcal{Z}(\eta)\delta\hat{\mu})^2+ (1+\mathcal{Z}(\eta)\hat{\mu}+\delta^2)^2}}\,.
\end{equation}
In Figure \ref{fig:imk}, we show the imaginary part of the Love number, $\Im(\tilde{k}_{2})$, as a function of liquid core radius $\eta=R_{\rm C}/R_{\rm P}$ for several different values of $\hat{\mu}=\mu/\rho g R_{\rm P}$ and of $\delta$.

It is evident that the runaway melting operates for a range of $\hat{\mu}$ and $\delta$. In Figure \ref{fig:instability}, we show the total tidal heating for a range of dimensionless rigidities and liquid core radii. These heating rates are calculated using Equations (\ref{eq:simpleheating}) and (\ref{eq:finalk2}), where we assume $\delta\ll 1$. We plot the quantity $\mathcal{E}(\eta)$ which we define as the ratio of the total heating as a function of liquid core radius to the unmelted case,
\begin{equation}\label{eq:epsilon}
\mathcal{E}(\eta)\equiv\frac{\dot{E}_{\rm Heat}(\eta)}{\dot{E}_{\rm Heat}(\eta=0)}\,. 
\end{equation}
Note that Figure 1 in \citet{Peale1979}, which shows the melting mechanism  for Io, corresponds to one horizontal slice of this plot. Note that the energy dissipated in the two layer case is greater than the energy in the homogeneous case for a wide range of the parameter space, roughly when $\hat{\mu}\gtrsim0.1$.

In most of the parameter space shown in Figure \ref{fig:instability}, the melting mechanism operates because the normalized total heating increases with liquid core radius. This occurs when the derivative of the heating with respect to the liquid core radius is positive, or when $d \mathcal{E}/ d \eta >0$. As a result, the runaway melting   saturates when this derivative is zero. In Figure \ref{fig:stability_analysis}, we show this derivative for the same range of liquid core radius and rigidity in Figure \ref{fig:instability}. Note that the contour of $d \mathcal{E}/d\eta=0$ indicates where the melting saturates and should correspond to the approximate location of the  liquid core radius. 

However, it is critical to note that this is only an approximate estimation of the liquid core radius which relies entirely on the model assumptions. More detailed previous calculations estimate a more accurate equilibrium by balancing heat production from mechanisms including both tidal and radiogenic heating against heat loss via mechanism such as heat piping, conduction and convection \citep{Barr2018,Dobos2019}. In this work, we  apply this idealized model to exoplanets for which  there are little near-term prospects for accurately measuring the core radii. Therefore, we have not performed a more detailed estimation of the equilibrium state incorporating additional --- yet unconstrained --- heat production and loss mechanisms. 

\section{Volcanism on Exoplanets}\label{sec:exocandidates} 

\begin{figure}
    \centering
    \includegraphics[width=1\linewidth]{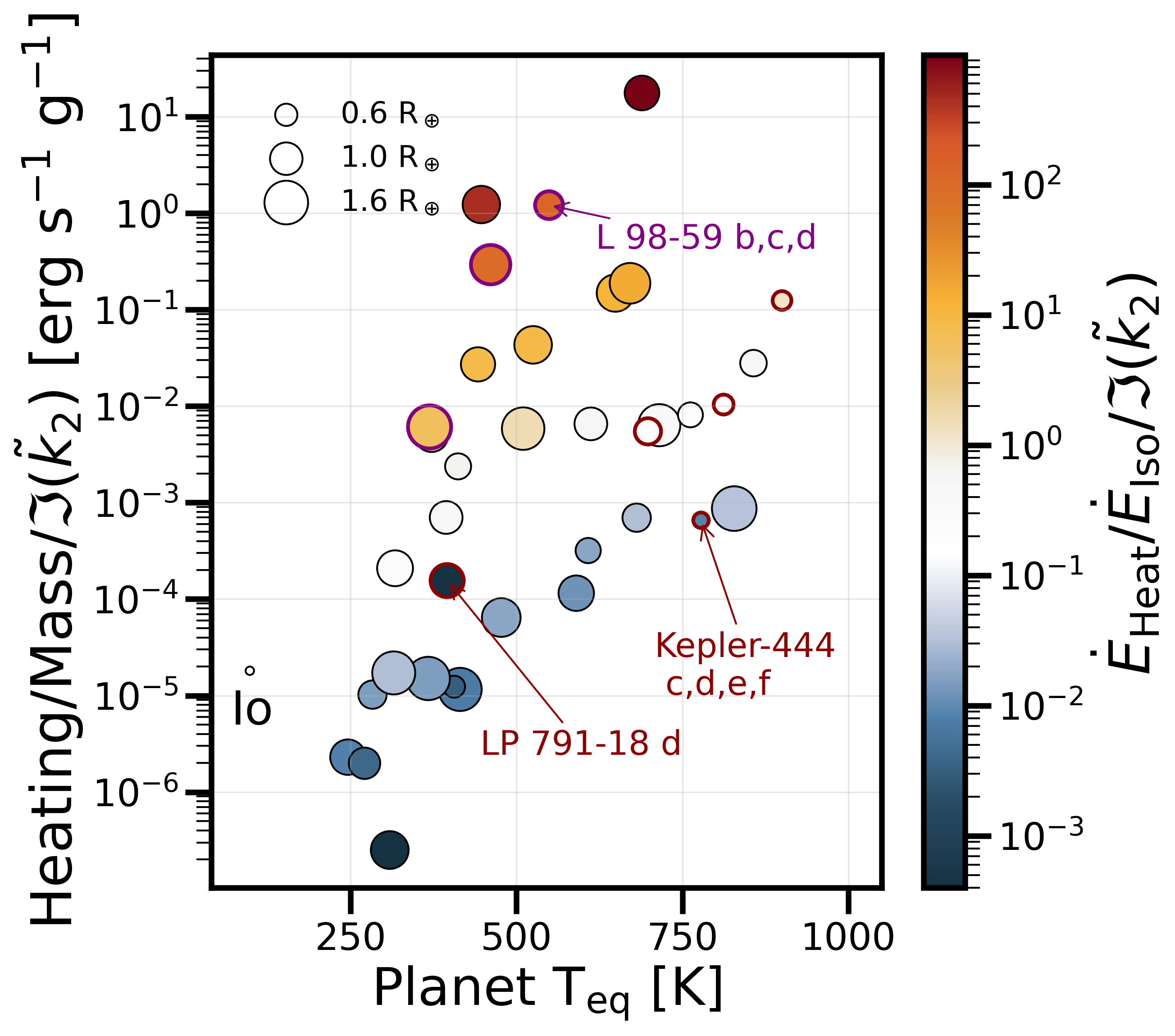}
    \caption{Tidal heating of exoplanets listed in Table \ref{table:candidates} normalized by the total mass, assuming a constant bulk density of $\rho=3$ g/cm$^3$. The color of each point indicates the total energy deposited via tidal heating (Equation (\ref{eq:scaled_heatingIO})) in comparison to the insolation heating from the host star (Equation (\ref{eq:insolation})). We only show planets with: (i) $R_{\rm P}\le1.6$ $R_\oplus$ (rocky), (ii)  effective planetary surface temperatures of $T_{\rm Eq}<1000$ K, (iii)  nonzero measured eccentricity and (iv) tidal heating comparable to or greater than that of Io (ignoring the uncertain quality factor or Love number).  The size of each point corresponds to the measured radius of the planet. } 
    \label{fig:candidates}
\end{figure}

\setlength{\tabcolsep}{1.5pt}
\begin{table*}[t]
\begin{center}
\caption{Exoplanets with properties that are amenable to the runaway melting and surface volcanism. The parameters shown are orbital period P$_{\rm Orb}$, eccentricity $e$, radius R$_{\rm P}$, planetary mass M$_{\rm P}$, planetary equilibrium temperature T$_{eq}$, stellar luminosity L$_*$, stellar mass M$_*$, effective temperature of the surface of the planet T$_\textrm{Eq}$, system distance $D$, J magnitude $J$,   tidal heating  $\dot{E}_{\rm Heat}/\Im(\tilde{k}_2)$, insolation heating $\dot{E}_{\rm Iso}$, transmission spectroscopy metric (TSM) \citep{Kempton2018} and the reference paper for the eccentricity value. Note that the value listed $\dot{E}_{\rm Heat}/\Im(\tilde{k}_2)$ corresponds to the case of $\Im(\tilde{k}_2)=1$. For reference the tidal heating of Io is $\simeq1.6\times10^{21}/\Im(\tilde{k}_2)$  erg s$^{-1}$. The table is organized by the planetary systems, so sets of planets within the same system are shown adjacent.  The reference for $e$ is provided in the last column, along with how the $e$ was derived (RV = radial velocity or radial velocity + transit fits; T = transit fits).}\label{table:candidates}
\begin{tabular}{ cccccccccccccc} 
Name & P$_\textrm{Orb}$ & $e$ & $R_{\rm P}$ & $M_{\rm P}$ & $T_\textrm{eq}$ & $L_\star$ & $M_\star$ & D & J & $\dot{E}_\textrm{Heat}/\Im(\tilde{k}_2)$ & $\dot{E}_\textrm{Iso}$ & TSM & $e$ Ref. \\
&[days]&&[R$_\oplus$] &[M$_\oplus$]&[K]&[L$_\odot$]&[M$_\odot$]&[pc]&[Mag ]&[erg s$^{-1}$]&[erg s$^{-1}$]&&\\
\hline
HD 136352 b & 11.58 & $0.079^{+0.068}_{-0.053}$ & 1.66 & 3.41 & 828.0 & 1.04 & 0.87 & 14.7 & 4.3 & 1.3$\times10^{25}$ & 3.7$\times10^{26}$ & 137 & RV; \cite{2020AJ....160..129K} \\
HD 219134 c & 6.77 & $0.062 \pm 0.039$ & 1.51 & 2.88 & 715.0 & 0.26 & 0.81 & 6.5 & 4.0 & 7.1$\times10^{25}$ & 1.7$\times10^{26}$ & 223 & RV; \cite{2017NatAs...1E..56G} \\
HD 219134 f & 22.72 & $0.148 \pm 0.047$ & 1.31 & 2.27 & 477.0 & 0.26 & 0.81 & 6.5 & 4.0 & 4.7$\times10^{23}$ & 2.5$\times10^{25}$ & 124 & RV; \cite{2017NatAs...1E..56G} \\
HD 23472 d & 3.98 & $0.069^{+0.053}_{-0.046}$ & 0.75 & 0.35 & 857.0 & 0.24 & 0.67 & 39.0 & 7.9 & 3.8$\times10^{25}$ & 8.6$\times10^{25}$ & 55 & RV; \cite{2022AA...665A.154B} \\
HD 23472 e & 7.91 & $0.056^{+0.054}_{-0.038}$ & 0.82 & 0.47 & 681.0 & 0.24 & 0.67 & 39.0 & 7.9 & 1.3$\times10^{24}$ & 4.1$\times10^{25}$ & 42 & RV; \cite{2022AA...665A.154B} \\
HD 23472 f & 12.16 & $0.048^{+0.044}_{-0.034}$ & 1.14 & 1.54 & 590.0 & 0.24 & 0.67 & 39.0 & 7.9 & 5.6$\times10^{23}$ & 4.4$\times10^{25}$ & 30 & RV; \cite{2022AA...665A.154B} \\
HD 260655 b & 2.77 & $0.039^{+0.043}_{-0.023}$ & 1.24 & 2.07 & 649.0 & 0.04 & 0.44 & 10.0 & 6.7 & 9.2$\times10^{26}$ & 7.7$\times10^{25}$ & 143 & RV; \cite{2022AA...664A.199L} \\
HD 260655 c & 5.71 & $0.038^{+0.036}_{-0.022}$ & 1.53 & 2.97 & 510.0 & 0.04 & 0.44 & 10.0 & 6.7 & 6.8$\times10^{25}$ & 4.5$\times10^{25}$ & 148 & RV; \cite{2022AA...664A.199L} \\
K2-116 b & 4.66 & $0.06_{-0.053}^{+0.142}$ & 0.69 & 0.26 & 762.0 & 0.18 & 0.69 & 49.4 & 8.6 & 8.7$\times10^{24}$ & 4.5$\times10^{25}$ & 41 & T; \cite{2017AJ....154..207D} \\
K2-128 b & 5.68 & $0.23^{+0.22}_{-0.19}$ & 1.42 & 2.61 & 671.0 & 0.15 & 0.71 & 114.6 & 10.5 & 1.8$\times10^{27}$ & 1.2$\times10^{26}$ & 10 & T; \cite{2017AJ....154..207D} \\
K2-129 b & 8.24 & $0.13_{-0.11}^{+0.21}$ & 1.04 & 1.12 & 372.0 & 0.01 & 0.36 & 27.8 & 9.7 & 1.8$\times10^{25}$ & 5.9$\times10^{24}$ & 33 & T; \cite{2017AJ....154..207D} \\
K2-136 b$^*$ & 7.98 & $0.14^{+0.12}_{-0.11}$ & 1.01 & 1.02 & 612.0 & 0.17 & 0.74 & 59.2 & 9.1 & 2.2$\times10^{25}$ & 4.1$\times10^{25}$ & 20 & RV; \cite{2023arXiv230402779M} \\
K2-136 d$^*$ & 25.58 & $0.071^{+0.063}_{-0.049}$ & 1.57 & 3.07 & 415.0 & 0.17 & 0.74 & 59.2 & 9.1 & 1.5$\times10^{23}$ & 2.1$\times10^{25}$ & 17 & RV; \cite{2023arXiv230402779M} \\
K2-266 c & 7.81 & $0.042_{-0.03}^{+0.043}$ & 0.7 & 0.28 & 608.0 & 0.15 & 0.69 & 77.6 & 9.6 & 3.6$\times10^{23}$ & 1.9$\times10^{25}$ & 19 & T; \cite{2018AJ....156..245R} \\
K2-72 b & 5.58 & $0.11_{-0.09}^{+0.20}$ & 1.08 & 1.28 & 442.0 & 0.01 & 0.27 & 66.4 & 11.7 & 1.1$\times10^{26}$ & 1.3$\times10^{25}$ & 18 & T; \cite{2017AJ....154..207D} \\
K2-72 c & 15.19 & $0.11_{-0.09}^{+0.20}$ & 1.16 & 1.65 & 317.0 & 0.01 & 0.27 & 66.4 & 11.7 & 1.1$\times10^{24}$ & 3.9$\times10^{24}$ & 13 & T; \cite{2017AJ....154..207D} \\
K2-72 d & 7.76 & $0.11_{-0.0.09}^{+0.21}$ & 0.73 & 0.31 & 412.0 & 0.01 & 0.22 & 66.4 & 11.7 & 3.0$\times10^{24}$ & 4.3$\times10^{24}$ & 44 & T; \cite{2017AJ....154..207D} \\
K2-72 e & 24.17 & $0.11_{-0.09}^{+0.20}$ & 0.82 & 0.48 & 283.0 & 0.01 & 0.22 & 66.4 & 11.7 & 1.8$\times10^{22}$ & 1.2$\times10^{24}$ & 28 & T; \cite{2017AJ....154..207D} \\
K2-91 b & 1.42 & $0.09^{+0.17}_{-0.07}$ & 1.1 & 1.37 & 689.0 & 0.01 & 0.29 & 62.6 & 11.4 & 7.6$\times10^{28}$ & 7.7$\times10^{25}$ & 44 & T; \cite{2017AJ....154..207D} \\
Kepler-138 b & 10.31 & $0.020 \pm 0.009$ & 0.59 & 0.15 & 406.0 & 0.04 & 0.51 & 66.9 & 10.3 & 8.3$\times10^{21}$ & 2.7$\times10^{24}$ & 20 & RV; \cite{2023NatAs...7..206P} \\
Kepler-138 c & 13.78 & $ 0.017^{+0.008}_{-0.007}$ & 1.58 & 3.13 & 367.0 & 0.04 & 0.52 & 66.9 & 10.3 & 1.9$\times10^{23}$ & 1.3$\times10^{25}$ & 16 & RV; \cite{2023NatAs...7..206P} \\
Kepler-138 d & 23.09 & $0.010 \pm 0.005$ & 1.26 & 2.13 & 309.0 & 0.04 & 0.52 & 66.9 & 10.3 & 1.6$\times10^{21}$ & 4.1$\times10^{24}$ & 10 & RV; \cite{2023NatAs...7..206P} \\
Kepler-444 c & 4.55 & $0.31^{+0.12}_{-0.15}$ & 0.5 & 0.08 & 900.0 & 0.37 & 0.76 & 36.4 & 7.2 & 5.1$\times10^{25}$ & 4.5$\times10^{25}$ & 88 & T; \cite{2015ApJ...799..170C} \\
Kepler-444 d & 6.19 & $0.18^{+0.16}_{-0.12}$ & 0.53 & 0.1 & 812.0 & 0.37 & 0.76 & 36.4 & 7.2 & 5.0$\times10^{24}$ & 3.4$\times10^{25}$ & 76 & T; \cite{2015ApJ...799..170C} \\
Kepler-444 e & 7.74 & $0.10^{+0.20}_{-0.07}$ & 0.42 & 0.04 & 778.0 & 0.37 & 0.62 & 36.4 & 7.2 & 1.6$\times10^{23}$ & 1.8$\times10^{25}$ & 135 & T; \cite{2015ApJ...799..170C} \\
Kepler-444 f & 9.74 & $0.29^{+0.20}_{-0.19}$ & 0.74 & 0.33 & 698.0 & 0.37 & 0.76 & 36.4 & 7.2 & 7.2$\times10^{24}$ & 3.7$\times10^{25}$ & 54 & T; \cite{2015ApJ...799..170C} \\
L 98-59 b & 2.25 & $0.103^{+0.117}_{-0.045}$ & 0.8 & 0.44 & 549.0 & 0.01 & 0.31 & 10.6 & 7.9 & 2.0$\times10^{27}$ & 1.7$\times10^{25}$ & 173 & RV; \cite{2021AA...653A..41D} \\
L 98-59 c & 3.69 & $0.103^{+0.045}_{-0.058}$ & 1.35 & 2.39 & 461.0 & 0.01 & 0.31 & 10.6 & 7.9 & 2.3$\times10^{27}$ & 2.4$\times10^{25}$ & 128 & RV; \cite{2021AA...653A..41D} \\
L 98-59 d & 7.45 & $0.074^{+0.057}_{-0.046}$ & 1.57 & 3.09 & 369.0 & 0.01 & 0.31 & 10.6 & 7.9 & 7.7$\times10^{25}$ & 1.3$\times10^{25}$ & 124 & RV; \cite{2021AA...653A..41D} \\
TOI-1266 c & 18.8 & $0.04 \pm 0.03$ & 1.56 & 3.06 & 315.0 & 0.03 & 0.45 & 36.0 & 9.7 & 2.1$\times10^{23}$ & 6.8$\times10^{24}$ & 25 & RV; \cite{2020AA...642A..49D} \\
TOI-2096 b & 3.12 & $0.15^{+0.27}_{-0.12}$ & 1.24 & 2.08 & 447.0 & 0.01 & 0.23 & 48.5 & 11.9 & 7.6$\times10^{27}$ & 1.7$\times10^{25}$ & 33 & RV; \cite{2023AA...672A..70P} \\
TOI-270 b & 3.36 & $0.034 \pm 0.025$ & 1.25 & 2.09 & 525.0 & 0.02 & 0.4 & 22.5 & 9.1 & 2.7$\times10^{26}$ & 3.0$\times10^{25}$ & 51 & RV; \cite{2021MNRAS.507.2154V} \\
TOI-700 b & 9.98 & $0.081^{+0.095}_{-0.058}$ & 1.01 & 1.01 & 394.0 & 0.02 & 0.42 & 31.1 & 9.5 & 2.4$\times10^{24}$ & 6.1$\times10^{24}$ & 29 & T; \cite{2020AJ....160..117R} \\
TOI-700 d & 37.42 & $0.111^{+0.140}_{-0.078}$ & 1.14 & 1.57 & 246.0 & 0.02 & 0.41 & 31.1 & 9.5 & 1.1$\times10^{22}$ & 1.4$\times10^{24}$ & 17 & T; \cite{2020AJ....160..117R} \\
TOI-700 e & 27.81 & $0.059^{+0.057}_{-0.042}$ & 0.95 & 0.82 & 271.0 & 0.02 & 0.41 & 31.1 & 9.5 & 5.6$\times10^{21}$ & 1.4$\times10^{24}$ & 21 & T; \cite{2023ApJ...944L..35G} \\
LP791-18 d &2.75  & $0.0015^{+0.00014}_{-0.00014}$ & 1.03&0.90&395.5&0.002 &  0.14  & 26.7 & 11.6 & $5.6 \times 10^{23}$ & $2.8 \times 10^{24}$ & 71 & RV; \cite{Peterson2023} \\

\end{tabular}
\end{center}
\end{table*}

\subsection{Exoplanets With Properties Amenable to Volcanism }
In this section, we calculate which currently-known terrestrial exoplanets could  exhibit volcanism driven by the melting mechanism outlined by \citet{Peale1978}.    \citet{Quick2020} calculated the  combined tidal and radiogenic  heating rates for 53 exoplanets with masses $M_{\rm P}<8 \,M_\oplus$ and  radii $R_{\rm P}<2 \,R_\oplus$. Our results generally agree with their findings, although we use different selection criteria. The main difference is that we only consider ``rocky'' planets, defined here to have radii $R_{\rm P}\lesssim \,1.6 R_\Earth$   \citep{Valencia2006,Valencia2007,Dorn2015,Fulton2017}. We also elected to \textit{not} include radiogenic heating in these calculations, the levels of which are unconstrained in exoplanets. For a more detailed discussion, we refer the reader to the third paragraph in \S \ref{sec:conclusions}.

To evaluate the importance of heating, we write the heating rate in the scaled form 
\begin{equation}\label{eq:scaled_heatingIO}
\begin{split}
     \dot{E}_\textrm{Heat} =\,\big(\, 3.4\times10^{25}\, \textrm{erg\,s}^{-1}\,\big)\,\,\bigg( \frac{P}{1\textrm{d}} \bigg)^{-5} \\\times\,\bigg( \frac{R_{\rm P}}{R_\oplus}\bigg)^5 \bigg(\frac{e}{10^{-2}}\bigg)^2\,\bigg(\frac{\Im(\tilde{k}_2)}{10^{-2}}\bigg)  \,.  
\end{split}
\end{equation}
We can compare this rate to that of insolation heating, which is given by
\begin{equation}\label{eq:insolation}
\begin{split}
    \dot{E}_\textrm{Iso}= \,\big(\,3.1\times10^{27}\, \textrm{erg\,s}^{-1}\,\big)\,\,\bigg(\frac{1-A}{1-A_\oplus}\bigg)\, \bigg(\frac{L_*}{L_\odot}\bigg) \\\,\times \,\,\bigg(\frac{M_*}{M_\odot}\bigg)^{-2/3}\,\bigg(\frac{R_{\rm P}}{R_\oplus}\bigg)^2\bigg(\frac{P}{1\textrm{d}}\bigg)^{-4/3} \,.
    \end{split}
\end{equation}
In Equation (\ref{eq:insolation}), $M_\odot$, $M_*$, $L_\odot$, and $L_*$ are the solar and stellar mass and luminosity respectively. The albedo of the planet is given by  $A$, and the equation is normalized to the albedo of the Earth, $A_\oplus$ which is approximately $A_\oplus=0.3$.

In addition to the radius constraint,  we only consider planets with finite eccentricity. Short-period planets are generally expected to have synchronized spin and zero obliquity because of the strong tidal torques. However, in the presence of misaligned companions, the spin of the planet may be captured into non-trivial ``Cassini-states" with finite obliquities --- such obliquites can be long-lasting even in the presence of tidal alignment torque \citep{Fabrycky2007,Millholland2019,Su2022}

As a final criteria, we conservatively only consider planets with surface temperatures less than 1000 K. Assuming that there is zero pressure at the surface, the condensation/evaporation temperatures of rocky, magnesium-silicate materials is $T\simeq 1300 {\rm K}$ \citep{Lodders2003,Sarafian2017}. 

In Figure \ref{fig:candidates} and Table \ref{table:candidates}, we show exoplanets that satisfy these criteria --- (i) rocky, (ii) nonzero eccentricity, and (iii) low surface temperature  ---  and may have active volcanism from runaway melting. The transmission spectroscopy metric (TSM) shown in Table \ref{table:candidates} was introduced by \citet{Kempton2018} (Equation (1) in that paper) and is a metric of the observability of an exoplanet atmosphere via transmission spectroscopy. The tidal heating of each planet is calculated using  Equation (\ref{eq:scaled_heatingIO}) and the mass is estimated with the mass radius relationship in \citet{Chen2017} for all cases. The equilibrium temperature is calculated assuming a constant Earth-like albedo for all cases. We also show the    ratio of the tidal heating (Equation (\ref{eq:scaled_heatingIO})) to the stellar heating (Equation (\ref{eq:insolation})). We normalize the heating parameters by $\Im(\tilde{k}_2)$, to indicate the uncertainty in the tidal properties of these bodies.  In Table \ref{table:candidates}, we show relevant orbital and planetary properties for all of these planets. 
 
Some of these planets have $\sim 5$ orders of magnitude more tidal heating per unit mass than Io --- assuming that the tidal quality factors are the same. However, even if the  quality factor (or Love number accordingly)  were orders of magnitude different from that of Io,  there may  be sufficient tidal heating to trigger melting --- or partial melting--- and volcanism in the most dramatic cases. Moreover, for the extreme cases, the tidal heating flux modulo the quality factor exceeds the instellation heating by a few orders of magnitude, or $\dot{E}_\textrm{Heat}/\Im(\tilde{k}_2)/\dot{E}_\textrm{Iso}\sim1-10^3$ (Figure \ref{fig:candidates}). Treating the tidal heating as an extra luminosity source, the equilibrium temperatures scale as $\dot{E}^{1/4}$.  For $\Im(\tilde{k}_2)\sim1-100$ the additional heating could be sufficient to raise the equilibrium temperatures by a factor of 1-5. For  L98-59 b where $\dot{E}_\textrm{Heat}/\Im(\tilde{k}_2)/\dot{E}_\textrm{Iso}\sim10^2$, assuming $\Im(\tilde{k}_2)\sim10$, the   equilibrium temperature is increased to $\sim1000$K.

Moreover, the inferred heating
rate per unit volume of radioactive nuclei in our solar system is $\sim0.006$ erg s$^{-1}$ cm$^{-3}$, which corresponds to a heating rate of $\dot{E}\sim 6.5 \times 10^{24}$ erg s$^{-1}$ \citep{Lodders2003}. This radioactive
heating per unit volume in the early Solar System is known to melt objects with sizes $\gtrsim$ 20 km \citep[Figure 4 and Equation (34) in][]{Adams2021}. Therefore exoplanets with $\dot{E}_\textrm{Heat}/\Im(\tilde{k}_2)\sim10^{27}$ erg s$^{-1}$ should experience  interior melting to some extent even if $\Im(\tilde{k}_2)\sim100$.

It is worth noting that the effective surface temperatures could be augmented by the tidal heating \citep{Becker2023}. However, even in the most dramatic case in Figure \ref{fig:candidates}, if $Q\sim10^2$, the tidal heating is still comparable to the stellar heating. It is also worth noting that we do not incorporate the melting  of the core in the mass estimate in Figure \ref{fig:candidates}. This choice is partially motivated  because the rigidity (which sets the liquid core radius as in Figure \ref{fig:stability_analysis}) is unconstrained. Moreover,  the total dissipation is typically not altered by an order of magnitude  due to  core melting (Figure \ref{fig:instability}).   The planet with the highest levels of heating (assuming constant tidal properties for all planets) is TOI-2096 b,  a super-Earth close to a 2:1 mean motion resonance with an exterior companion \citep{Pozuelos2023}.

A promising  system for exhibiting volcanism is the L 98-59  multi-planet system \citep{Kostov2019,Demangeon2021}. This small M-dwarf harbours four planets, and the outer three planets are close to a $2:4:7$ mean motion resonance, reminiscent of the  Laplace resonance in the Galilean satellites. It is possible that the  nonzero eccentricities of the planets  --- despite the significant tidal heating --- could be explained by forced eccentricity from the mean motion resonance. Measurements obtained with the \textit{Hubble Space Telescope (HST)}  of  L 98-59 b were consistent with no atmosphere \citep{Damiano2022,Zhou2022}. However, \citet{Barclay2023} presented preliminary and marginal evidence  of an atmosphere on  L 98-59 c with \textit{HST}  measurements,  although this has since been debated \citep{Zhou2023}.

Another intriguing system is Kepler-444 \citep{Campante2015}, an old triple star system whose main star hosts five small rocky exoplanets  \citep{Winters2022}. However, \citet{Stalport2022} performed a dynamical analysis of the system and found that none of the planets are in low order 2 or 3 planet mean motion resonances.

\subsection{Dynamical Constraints on Tidal Properties}

It is tempting to replicate the analysis performed by \citet{Peale1979} for Io with these planets that appear to have the prerequisites for volcanism and that are in mean motion resonances. \citet{Peale1979} estimated the quality factor of Io by equating the increase in Io's mean motion from tidal dissipation (Equation 8 in that work) to the decrease in mean motion from tidal transfer from Jupiter's rotation (Equation 9 in that work). However, in the Jupiter Io case, the  rotation period of Jupiter is \textit{faster} than the orbital period of Io so the two tidal forces act in opposite directions. However,  the stellar rotation rates are likely slower than the orbital periods of the planets.  Therefore, the  tidal forces will operate coherently and  --- somewhat disappointingly ---  the quality factor of these planets cannot be estimated in a similar fashion.  

It is possible that the eccentricities of these planets are maintained  via the mean motion resonances. For example, the planet LP 791-18d has nonzero forced eccentricity, which led \citet{Peterson2023} to claim that it would have active volcanism. Unfortunately, the value of the forced eccentricity  does not provide useful constraints on the tidal properties.  To illustrate this, Equation (17) in \citet{Lithwick2012} can be solved --- by noting that $|z_1\simeq\mu n_2/(\gamma^2+(2 n_2 \Delta)^2)^{1/2}\le \mu n_2/\gamma|$ using the notation in that paper and rearranging--- to yield the following constraint for planets in mean motion resonances:

\begin{equation}\label{eq:LithwickWu}
\begin{split}
    \gamma_e \le \frac{ M_1}{ M_*} \frac{n_2}{e_1} \simeq (0.08 \,\textrm{yr}^{-1})\,\bigg(\frac{M_1}{2.2 M_\oplus}\bigg)\\\times\bigg(\frac{0.27 M_\odot}{M_*}\bigg)\bigg(\frac{P_2}{7.45 \text{ d}}\bigg)\bigg(\frac{0.1}{e_1}\bigg)\,.
    \end{split}
\end{equation}
In Equation (\ref{eq:LithwickWu}) $\gamma_e$ is the damping rate of eccentricity for a pseudo-synchronized planet $\gamma_e = (1/e)\, \textrm{d}e/\textrm{dt}$ and the subscripts indicate the inner and outer of two planets in a mean motion resonance. On the right hand side of Equation (\ref{eq:LithwickWu}), we have substituted values for L 98-59 c and d (which are close to a 2:1 resonance). Typical values of the damping rates from tidal dissipation are much smaller than this, and for L98-59c as an example with planetary parameters from \citet{2021AA...653A..41D}, 

\begin{equation}\label{eq:gammae}
\begin{split}
    \gamma_e = -\frac{21}{2} \Im(\tilde{k}_2) \, n\,\bigg(\,\frac{M_*}{M_{\rm P}} \,\bigg)\, \bigg(\,\frac{R_{\rm P}}{a}\,\bigg)^5  \\
    \simeq  (6.8\times 10^{-6} \,\textrm{yr}^{-1})\ \Im(\tilde{k}_2) \bigg(\frac{2.4 M_\oplus}{M_{\rm P}}\bigg)\\ \times\bigg(\frac{M_*}{0.31 M_\odot}\bigg) \bigg(\frac{R_{\rm P}}{1.35 R_\oplus}\bigg)^5 \,\\\times\bigg(\frac{0.0304 \,\textrm{au}}{a}\bigg)^5\,\bigg(\frac{3.69 \,\textrm{d}}{P}\bigg)\,.
\end{split}
\end{equation}
Unfortunately, this   does  not provide a useful constraint on the tidal properties.

\begin{figure}
    \centering
    \includegraphics[width=1\linewidth]{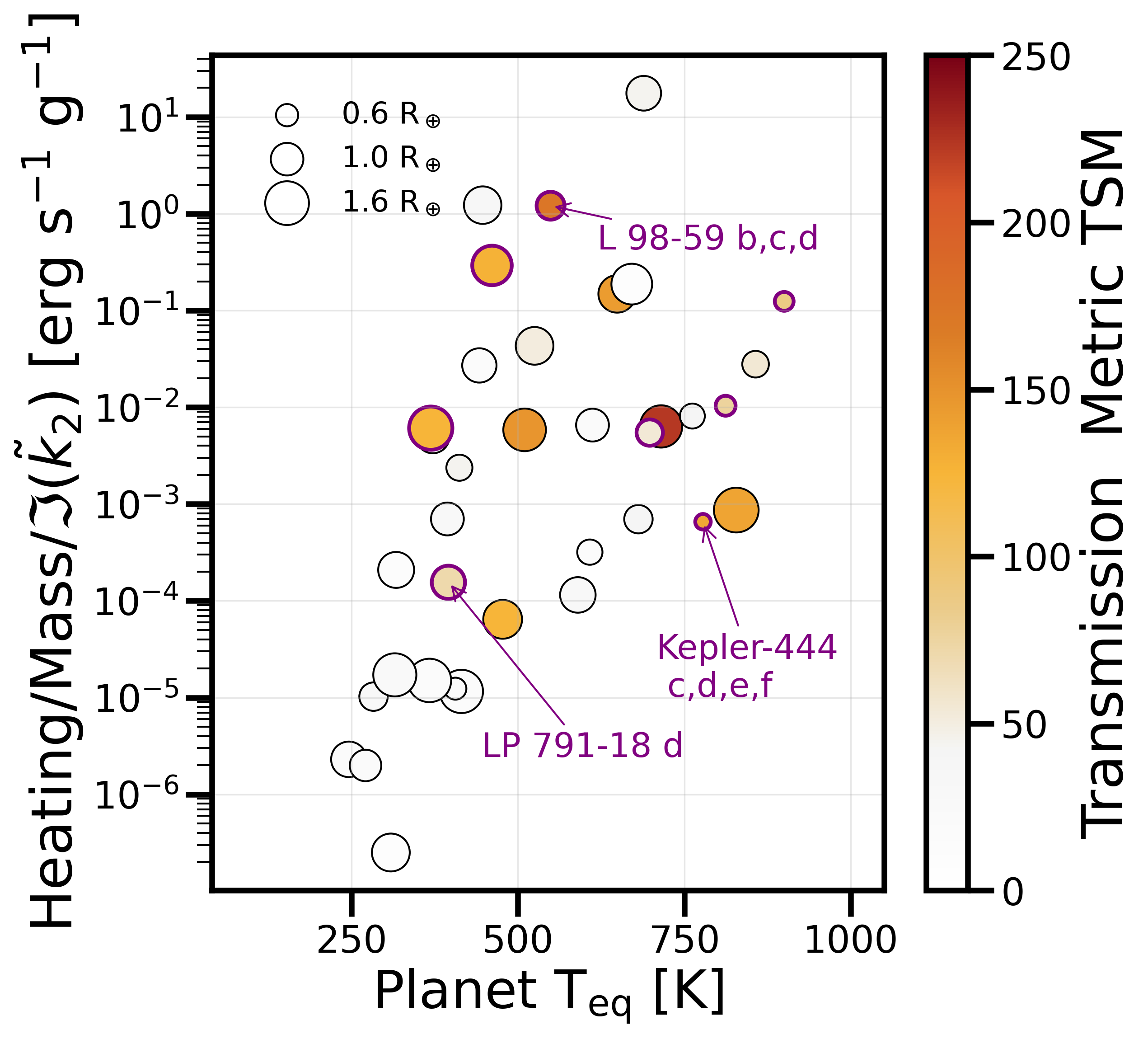}
    \caption{Same as Figure \ref{fig:candidates} except the color of each point corresponds to the value of the  transmission spectroscopy metric given by \citet{Kempton2018}.  } 
    \label{fig:candidatesTSM}
\end{figure}

\section{Observability with JWST}\label{sec:future_prospect}

In this section, we investigate the feasibility of detecting volcanism on exoplanets with particular attention to the L98-59 planets as a case study. Volcanic outflows on Earth typically consist of H$_2$O, CO$_2$, SO$_2$, H$_2$S and H$_2$ \citep{Halmer2002}. \citet{Kaltenegger2010} demonstrated that SO$_2$ is a promising tracer of exoplanetary volcanism, given its high abundance from Earth's volcanic activity and that it is not typically in the atmospheres of normal planets. SO$_2$ has a variety of spectral features detectable in the infrared, for example with \textit{JWST}, (2.5, 4.05, 7.34, 8.68, and 19.30 \um) as does H$_2$S  (3.80 and 8.45 micron) \citep{Kaltenegger2010,Kaltenegger2010b,Fortin2022}. In \ref{subsec:JWST} we focus on the detectability of \sulfurdioxide. 

\sulfurdioxide\ has recently been observed in the atmosphere of the gas giant planet WASP-39\,b \citep{alderson23, rustamkulov23} as part of the \textit{JWST} Transiting Exoplanet Early Release Science program \citep{stevenson16, bean18, ahrer23, feinstein23}. However, there have been no published detections of \sulfurdioxide\ for a rocky planet. Motivated by the possibility of detecting SO$_2$ via transmission spectroscopy, we show a metric of the observability of the atmospheres of these potentially volcanic exoplanets in Figure \ref{fig:candidatesTSM}. The color indicates the value of the transmission spectroscopy metric introduced by \citet{Kempton2018} (Equation (1) in that paper). 

It appears that the L98-59 system is not only one of the most promising systems for planetary volcanism but is also is the most viable system for follow-up observations with transmission spectroscopy with \textit{JWST}. In the following subsection, we provide modeled spectra for the L98-59 system assuming different amounts of \sulfurdioxide\ in each planet's atmosphere. 

\subsection{Atmospheric Model Description}

We generate atmospheric models of the three planets in the L98-59 system using Aurora \citep{Welbanks2021}. Aurora numerically computes the line-by-line radiative transfer in transmission geometry assuming hydrostatic equilibrium. We compute these models using 100 layers uniformly spaced in $\log_{10}$(P) between $10^2$ bar and $10^{-9}$~bar, at a constant resolution of R=10$^5$ between $0.5\,\mu$m and $15\,\mu$m. We consider three main atmospheric scenarios, ranging from CO$_2$- to SO$_2$-rich, with trace amounts of H$_2$O, H, and He (see Table~\ref{tab:comps}). For these chemical species, we use line lists from \citet{Rothman2010} for H$_2$O and CO$_2$, \citet{Underwood2016} for SO$_2$, and \cite{Richard2012} for H$_2$-H$_2$ and H$_2$-He collision induced absorption (CIA, mostly negligible in the chosen scenarios). These initial models consider cloud-free atmospheres only, as we aim to inform the observational prospects of these optimistic scenarios. Future investigations will consider the effects of different cloud and haze species, as well as other chemical scenarios.  

We consider three atmospheric scenarios summarized in Table \ref{tab:comps}. First, we consider a CO$_2$-rich atmosphere composed of 98\% CO$_2$, 5\% SO$_2$, 1\% H$_2$O, and a mixture of H$_2$ and He in the remaining percentage, assuming a He/H$_2$ solar ratio of 0.17 \citep{Asplund2009}. Second, we consider a scenario where the atmosphere is composed of 50\% SO$_2$, 48\% CO$_2$, 1\% H$_2$O, and a 1\% solar mixture of H$_2$ and He. Finally, we consider a scenario of SO$_2$ dominated with a 98\% SO$_2$ composition with 1\% H$_2$O and 1\% H$_2$ and He. The first scenario aims to assess the observability of a composition loosely similar to a Venus-like CO$_2$-rich atmosphere, while the third scenario considers a scenario loosely similar to Io, with an SO$_2$-rich atmosphere. The second scenario is an intermediate case between these two extremes.

\begin{deluxetable}{c c c c c}[!ht]
\tabletypesize{\footnotesize}
\tablecaption{Summary of Atmospheric Compositions \label{tab:comps}}
\tablehead{\colhead{Scenario} & \colhead{H$_2$ + He [\%]} & \colhead{CO$_2$ [\%]} & \colhead{H$_2$O [\%]} & \colhead{\sulfurdioxide [\%]}\\
\hline
1 & 1 & 93 & 1 & 5\\
2 & 1 & 48 & 1 & 50\\
3 & 1 & 0 & 1 & 98
}
\startdata
\enddata
\end{deluxetable}

\begin{figure}
    \centering
    \includegraphics[width=1\linewidth]{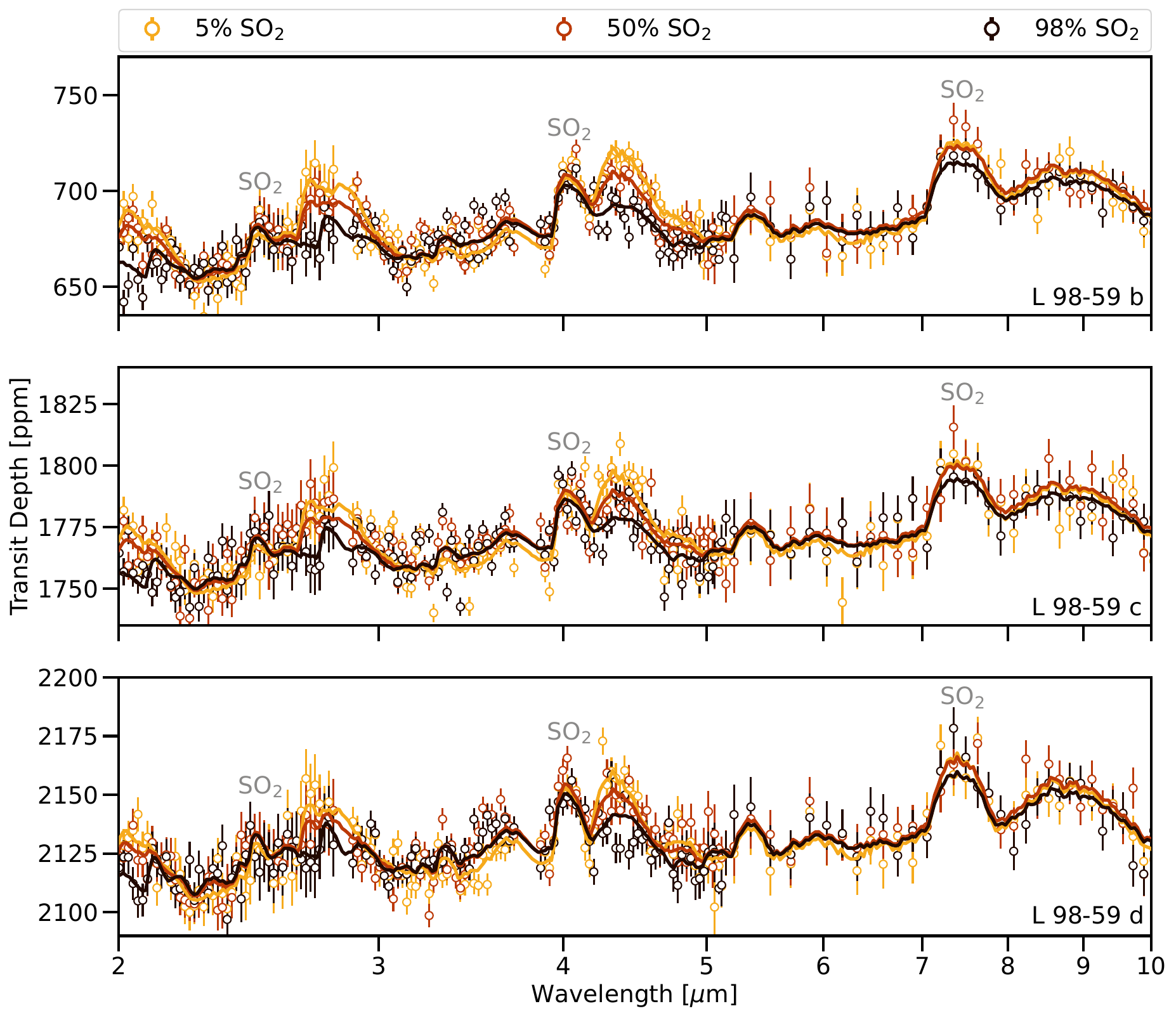}
    \caption{Modeled transmission spectra for L98-59\,bcd. We simulated observations using NIRISS/SOSS ($\lambda = 0.8 - 2.8$\,\um; $R \sim 150$), NIRSPec/G395H ($\lambda = 3-5$\,\um; $R \sim 150$), and MIRI/LRS ($5 - 10$\,\um; $R \sim 80$) to determine the SNR for each of the three \sulfurdioxide\ absorption features at 2.5, 4.05, and 7.34\,\um. These spectra represent the results of 5 transits per each target. The models are overplotted as solid lines.} 
    \label{fig:spectra}
\end{figure}
\begin{figure*}
    \centering
    \includegraphics[width=1\linewidth]{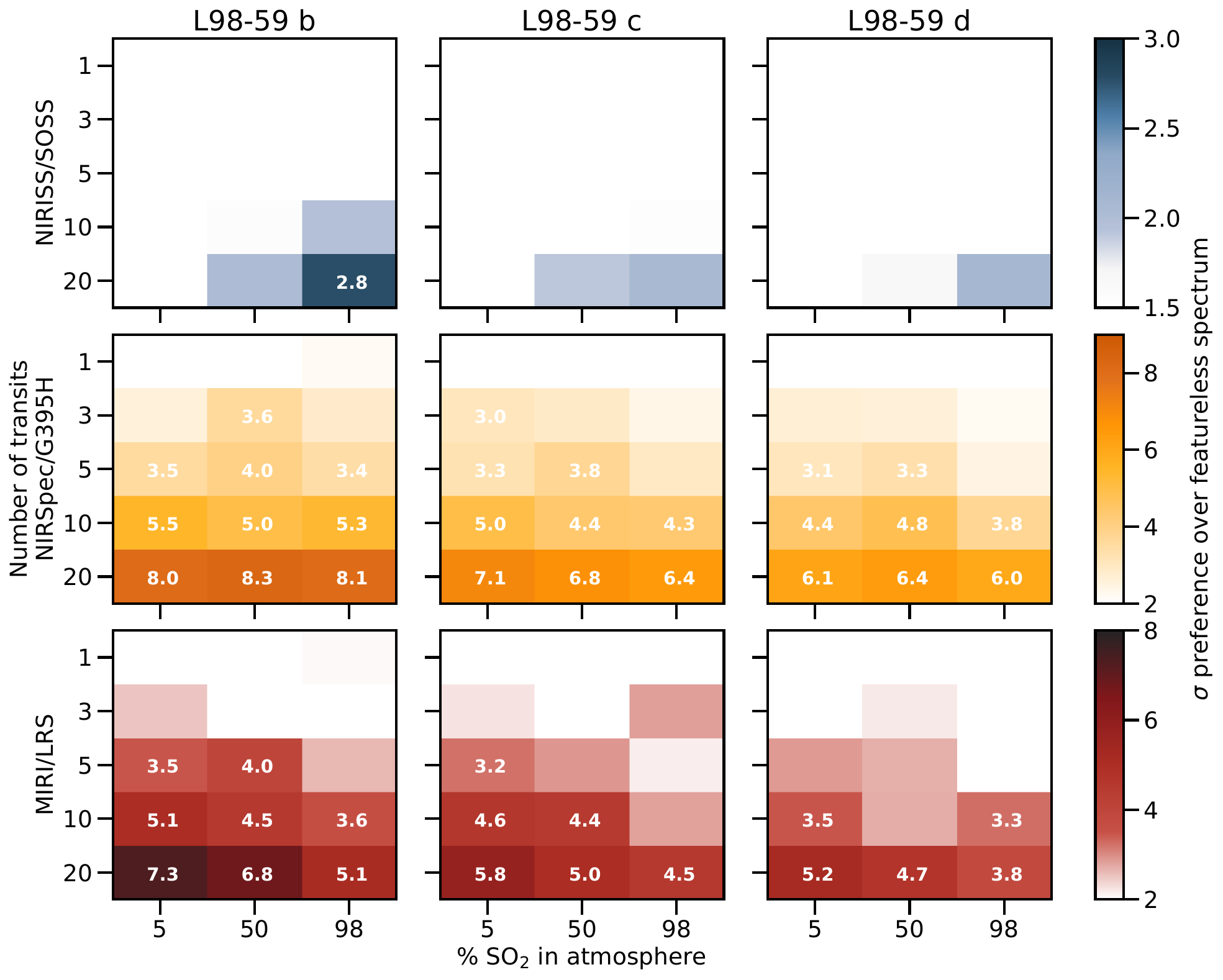}
    \caption{$\langle SNR \rangle$ preference of an atmosphere with \sulfurdioxide\ over a featureless transmission spectrum (flat line) for L~98-59\,bcd as a function of number of transits observed. For any significant ($> 3\sigma$; labeled) detection of \sulfurdioxide\ in the atmosphere of these three planets, we would need to observe at least $\sim$10~transits with either NIRSpec/G395H or MIRI/LRS. The \sulfurdioxide\ feature at $\lambda = 2.5$\,\um\ would require observations of $> 20$~transits for any significant detection, given the relatively weaker absorption at the shorter wavelengths. It is notable that   the significance of detection decreases with the \sulfurdioxide\ fraction. } 
    \label{fig:so2_grid}
\end{figure*} 

\subsection{Synthetic JWST Observations}\label{subsec:JWST}

Within the wavelength coverage of \textit{JWST}, \sulfurdioxide\ has absorption features at 2.5, 4.05, 7.34, and 8.68\,\um. We use the aforementioned models for each of the L98-59 planets and simulate  observations of their atmospheres with NIRISS/SOSS, NIRSPec/G395H, and MIRI/LRS. To do this, we run our models through \texttt{pandexo 2.0}, an open-source Python package for generating instrument simulations for  \textit{HST}  and \textit{JWST} \citep{pandexo}. We assumed the transit duration, $T_{14}$, presented in \cite{Demangeon2021} and assumed an out-of-transit duration of $3 \times T_{14}$. The resulting simulated transmission spectra are presented in Figure~\ref{fig:spectra}.

The \sulfurdioxide\ absorption features at $\lambda = 4.05, 7.34$\,\um\ are visible by-eye in each of the simulated spectra, while the \sulfurdioxide\ absorption feature at $\lambda = 2.5, 8.68$\,\um\ is likely to be overwhelmed by either H$_2$O absorption or instrument noise. We find a maximum $\sim 10$\,ppm difference between the 5\% and 98\% \sulfurdioxide\ atmospheres at $\lambda = 4.05, 7.34$\,\um\ for all three planets. While the  depth of the \sulfurdioxide\ does not change significantly across all three scenarios, the depths of other molecules, such as CO$_2$, do. Therefore, it may be possible to distinguish these atmospheric compositions via broad wavelength observations, which will allow for accurate measurements of the abundances of other species.

We determine the detectability of each \sulfurdioxide\ absorption feature as a function of the number of transits to observe,  $n$. We use Equation (5) of \citet{lustigyaeger19} in order to calculate the  expected signal-to-noise (SNR) of each feature, which is given by

\begin{equation}\label{eq:SNR}
    \langle \,  \textrm{SNR} \, \rangle = \sqrt{\sum_{i=1}^{N_\lambda} \left( \frac{m_{1,i} - m_{2,i}}{\sigma_i} \right)^2} \,.
\end{equation}
In Equation (\ref{eq:SNR}),  $m_1$ is the transmission model, $m_2$ is a featureless transmission model, and $\sigma$ is the error on the observations. The subscript $i$ for each of these parameters indicates that the residual is computed at  discrete resolution elements and ${N_\lambda}$ is the total number of resolution elements. We define the featureless transmission model as the best-fitting constant planet radius with wavelength. We evaluate the $\langle \textrm{SNR} \rangle$ as $\lambda = [2.45, 2.55], [3.92, 4.20]$, and $[7.0, 7.9]$, which broadly covers all three \sulfurdioxide\ features. We repeat this calculation for $n = 1, 2, 5, 10, 20$ transits observed with each \textit{JWST} instrument. The results for L98-59\,bcd with NIRISS/SOSS, NIRSpec/G395H, and MIRI/LRS are presented in Figure~\ref{fig:so2_grid}.

Using the metric above, we find that a significant (i.e., $>3\sigma$) preference for an atmosphere with \sulfurdioxide\ over a featureless spectrum (``flat line") with NIRISS/SOSS in under 20 transits observed is unlikely. On the other hand, we find that it is likely to prefer an atmosphere with some amount of \sulfurdioxide\ for all three L98-59 planets in 3-5 transits if observed with NIRSpec/G395H. It is possible to prefer an atmospheric model with \sulfurdioxide\ in just three transits of L98-59\,c with NIRSpec/G395H. In order to detect an atmospheric composed of 98\% \sulfurdioxide\ for L98-59\,bcd, 5-10 transits would need to be observed. While there is more \sulfurdioxide\ in the atmosphere, this results in an atmosphere with a higher mean molecular weight, and therefore the absorption features are more muted \citep{dewit13}.

Preferring \sulfurdioxide\ over a featureless spectrum with MIRI/LRS may be feasible with 5 - 10 transits for L98-59\,bc, and 10-20 transits for L98-59\,d. Observing multiple \sulfurdioxide\ features with different instruments would provide the greatest confidence in the planetary nature of the absorption feature \citep[Dyrek et al. submitted; Powell et al. submitted;][]{alderson23, rustamkulov23}. Furthermore, broad wavelength coverage would allow us to better constrain the abundances of other species such as H$_2$O and CO$_2$ and will  provide insights into the relative abundance of \sulfurdioxide. Assessing the detection of a chemical species within a Bayesian framework would require performing atmospheric retrievals and an in-depth model comparison and criticism \citep[see e.g.,][for a discusssion]{Welbanks2023}. Such exploration with atmospheric retrievals would enable an assessment of the ability to infer the abundance of \sulfurdioxide and other trace species. This exploration is beyond the scope of this paper and we reserve it for a follow-up study.

Five transits of L~98-59\,b, one transit of L 98-59\,c, and two transits of L 98-59\,d will be observed with NIRSpec/G395H during \textit{JWST} Cycles 1 and 2 (GO 2521, 3942, 4098; GTO 1201). The observations of L 98-59\,cd will likely be insufficient for any significant detection of \sulfurdioxide; however, if the atmosphere of L 98-59\,b  has any amount of  \sulfurdioxide, it is possible these observations would be able to detect the absorption feature at 4.05\,\um at $\geq 3\sigma$. 
More transits of L 98-59\,cd would need to be observed to search for evidence of a \sulfurdioxide-rich atmosphere.

\section{Discussion}\label{sec:conclusions}

In this paper we have generalized the runaway melting mechanism proposed by \citet{Peale1978} and \citet{Peale1979}  to extrasolar planets. We then identified candidate exoplanets for which this mechanism may be inducing surface volcanic activity detectable with \textit{JWST}. It appears that the L 98-59 system is the most promising system to detect volcanism with a reasonable number of transits with \textit{JWST} observations. The detection of volcanism on extrasolar planets  could be used to constrain their interior structure and tidal properties. 

Although the detectability of volcanic activity on exoplanets is promising, a number of caveats to the current analysis must be discussed.  We have ignored the effect of planet-planet tides, which can be induced in short period, multi-planet systems and which would complicate the scenario. The nonzero eccentricities, which may be caused by mean motion resonances, provide an effective mechanism to sustain long-term tidal heating (\citealt{Henning2009}). However, \citet{HayMatsuyama2019} concluded that stellar tides consistently dominate over planet-planet tides for a range of material viscosities \citep[see also][]{Zanazzi2019}.

 The melting mechanism also only operates if  core melting is initiated via another process, such as radiogenic heating --- the primary source of internal heating on the Earth \citep{Davies1999}.  The origin of radiogenic species such as $^{26}$Al and $^{10}$Be in the early Solar System is unclear, and it is debated still if they formed in situ via spallation \citep{Shu1996,Gounelle2001,Leya2003,Duprat2007,adams2010} or were delivered from supernovae enrichment \citep{Goswami2005,Krot2009,Cameron1977,Hester2004,adams2010}.  It is currently unclear to what extent these exoplanets would experience internal melting from radiogenic heating, as the initial planetary abundances of radioactive nuclei may vary significantly depending on the formation location with in the protoplanetary disk  \citep{Cleeves2013, Adams2021}. A variety of stochastic processes may lead to significant abundance variations in radiogenic heating rates \citep[by factors of 10-20 or greater;][]{Adams2021} between planets and between exoplanet systems \citep{Curry2022}. The exoplanets considered in this paper that are most visible with \textit{JWST} in terms of the TSM have radial locations near the initial disk truncation radius which is also where the spallation effects are expected to be 
the largest \citep{Adams2021}.

In \S\ref{sec:future_prospect}, we focused on IR observations that would indicate the presence of SO$_2$ in the atmosphere of these planets. Although SO$_2$ could be produced by volcanic activity, it is also possible that it could be caused by other processes. A promising alternative avenue  to infer exoplanetary volcanism is by detecting  plasma tori  injected by volcanic activity \citep{Kislyakova2018,Kislyakova2019}. This structure would be analogous to the Jovian plasma torus, which sources plasma predominantly from Io's volcanically produced SO$_2$ (see the discussion in Section \ref{sec:intro}). While we focused on the  observability of SO$_2$ with \textit{JWST} in this work, detections of plasma torus features in addition to SO$_2$ would strengthen a claim of exoplanetary volcanism.

\section{Acknowledgements}
We thank   Sam Cabot,  Leslie Rogers,  Nikole Lewis, Samantha Trumbo, Jonathan Lunine, Ray Jayawardhana, L\'igia Fonseca Coelho, Jonas Biren, Christopher O'Connor, JT Laune, Greg Laughlin,  Rafael Luque, Garrett Levine, Charles Cockell,  Jacob Bean, Ngoc Truong, Abby Boehm and  Suvrath Mahadevan  for useful conversations and suggestions. We thank the anonymous reviewer for extremely insightful feedback that greatly strengthened the content of this manuscript.

 D.Z.S. acknowledges financial support from the National Science Foundation  Grant No. AST-2107796, NASA Grant No. 80NSSC19K0444 and NASA Contract  NNX17AL71A. D.Z.S. is supported by an NSF Astronomy and Astrophysics Postdoctoral Fellowship under award AST-2202135. This research award is partially funded by a generous gift of Charles Simonyi to the NSF Division of Astronomical Sciences.  The award is made in recognition of significant contributions to Rubin Observatory’s Legacy Survey of Space and Time. ADF acknowledges support by the National Science Foundation Graduate Research Fellowship Program under Grant No. (DGE-1746045) and funding from NASA through the NASA Hubble Fellowship grant HST-HF2-51530.001-A awarded by STScI. A.G.T. acknowledges support by the Fannie and John Hertz Foundation and the University of Michigan's Rackham Merit Fellowship Program. This research has made use of the NASA Exoplanet Archive, which is operated by the California Institute of Technology, under contract with the National Aeronautics and Space Administration (NASA) under the Exoplanet Exploration Program.

\software{\texttt{pandexo 2.0} \citep{pandexo}}

\bibliography{bibs}{}
\bibliographystyle{aasjournal}

\end{document}